# KA-Band Mobile Antenna for Satellite Communication

*Sidra Tul. Muntaha, Ahmad Arfeen, Kashan Raza*

***Abstract*— This research focuses on the design of Ka-band mobile antennas for satellite communication operating at 29 GHz. Starting from a single element and progressing to an 8x8 array, the antennas achieved a gain of up to 21 dB and return losses as low as -30 dB. The design process involves mathematical calculations and software implementation, utilizing parameters like patch dimensions, substrate properties, and effective permittivity.**

**The chosen Ka-band frequency range, known for higher data transfer rates, addresses the demand for swift communication. Challenges in Ka-band mobile antenna design, including signal attenuation, directional accuracy, circular polarization, and impedance matching, are addressed through various configurations, including phased-array and electronically steerable antennas.**

**This research focuses on the design of Ka-band mobile antennas for satellite communication at 29GHz, progressing from a single element to an 8x8 array. The antennas achieved a gain of up to 21 dB and return losses as low as -30 dB through mathematical calculations and software implementation using CST. Challenges in Ka-band antenna design, such as signal attenuation and impedance matching, are addressed through various configurations, including phased-array and electronically steerable antennas. Integration of machine learning techniques aids in optimization. In conclusion, this research advances high-frequency transmission technology, meeting the demands of modern satellite-based communication systems for applications like high-speed internet access and multimedia streaming.**

## I. INTRODUCTION

The integration of Ka-band mobile antennas into satellite communication has marked a significant stride in advancing high-frequency transmission technology. Operating within the frequency range of 26.5 to 40 gigahertz, the Ka-band introduces enhanced data transfer rates and expanded bandwidth, strategically addressing the escalating demand for swift and efficient communication in modern satellite-based systems. This research focuses on the design and optimization of Ka-band mobile antennas, specifically operating at 28 GHz, with the goal of achieving optimal performance.

Mobile antennas play a pivotal role in facilitating the transmission and reception of signals over various frequency bands. Their compact and versatile nature makes them ideal for integration into mobile devices, amplifying connectivity and supporting applications such as high-speed internet access and multimedia streaming. The seamless exchange of voice, data, and multimedia content is crucial in enhancing the coverage, reliability, and overall performance of wireless communication networks.

Efficient antenna placement becomes paramount in maximizing signal coverage and minimizing interference, especially considering the challenges posed by size constraints and interference. The energy efficiency and power consumption of antennas are crucial factors; particularly as mobile devices operate on limited battery power. Thus, striking a balance between strong signals and minimal energy consumption becomes essential for the sustainability of mobile communication.

The physical size of antennas is a key consideration due to the compact nature of mobile devices. Engineers focus on miniaturizing antennas while maintaining or improving performance to meet the demand for sleek and portable mobile devices without compromising on communication capabilities.

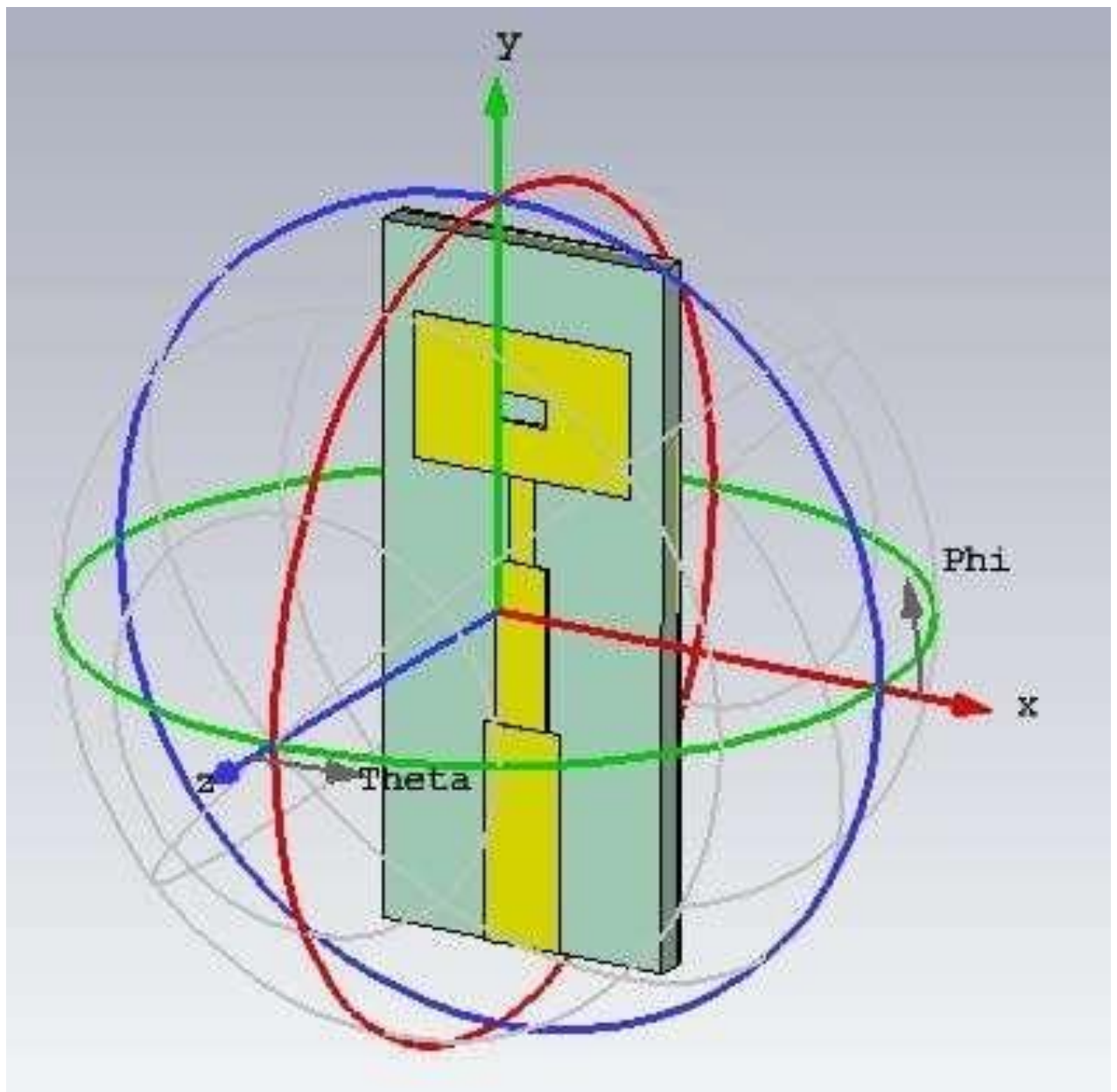

*Figure 1-antenna design*

Satellite communication, a cornerstone of our modern interconnected world, extends its reach to remote areas,

providing seamless connectivity by overcoming the limitations of traditional terrestrial networks. In this context, the collaboration between mobile devices and satellites plays a crucial role, enabling direct communication and global coverage. The research explores various satellite orbits, including Geostationary Earth Orbit (GEO), Medium Earth Orbit (MEO), and Low Earth Orbit (LEO), each offering unique advantages for specific applications.

The Ka-band, with its higher frequencies and improved spectrum efficiency, emerges as a key player in satellite communication. Its ability to support increased data rates makes it an ideal choice for bandwidth-intensive applications such as video streaming and large data transfers. However, challenges persist in designing efficient Ka-band mobile antennas, including signal attenuation and the need for directional accuracy, circular polarization, and impedance matching. Researchers have explored innovative antenna configurations, including phased-array and electronically steerable antennas, to address these challenges.

The integration of machine learning techniques, such as artificial intelligence algorithms, introduces a new dimension to the optimization of Ka-band mobile antennas. By analyzing vast datasets generated through electromagnetic simulations, these algorithms contribute to identifying optimal antenna configurations, minimizing signal losses, and enhancing overall performance.

The research methodology involves mathematical calculations and software implementation using tools like CST and MATLAB to design and optimize micro strip patch antennas for Ka-band satellite communication. The goal is to achieve high performance, with a gain of up to 21 dB and return losses as low as -30 dB, particularly in an 8x8 array configuration.

## II. ANTENNA DESIGN

This section details the theoretical parameters and design specifications for a single-element microstrip patch antenna tailored for Ka-Band satellite communication. Employing a microstrip edge feed configuration and Rogers RT Duroid 5880 as the dielectric substrate with a precise dielectric constant of 1.96 ± 0.02, the antenna is designed to operate optimally at a height of 0.784 mm. The compact dimensions of 3.08x4.08 mm are meticulously chosen to meet the requirements of high-frequency transmission.

The antenna's theoretical parameters, including gain, return loss, polarization, and resonating frequency, are crucial for optimizing performance in the Ka-band frequency range. Utilizing CST simulation software, the antenna undergoes comprehensive analysis to ensure its efficiency and effectiveness in satellite communication systems. The presented parameters serve as a foundational framework for the practical implementation and experimental validation of the single-element microstrip patch antenna in Ka-band applications.

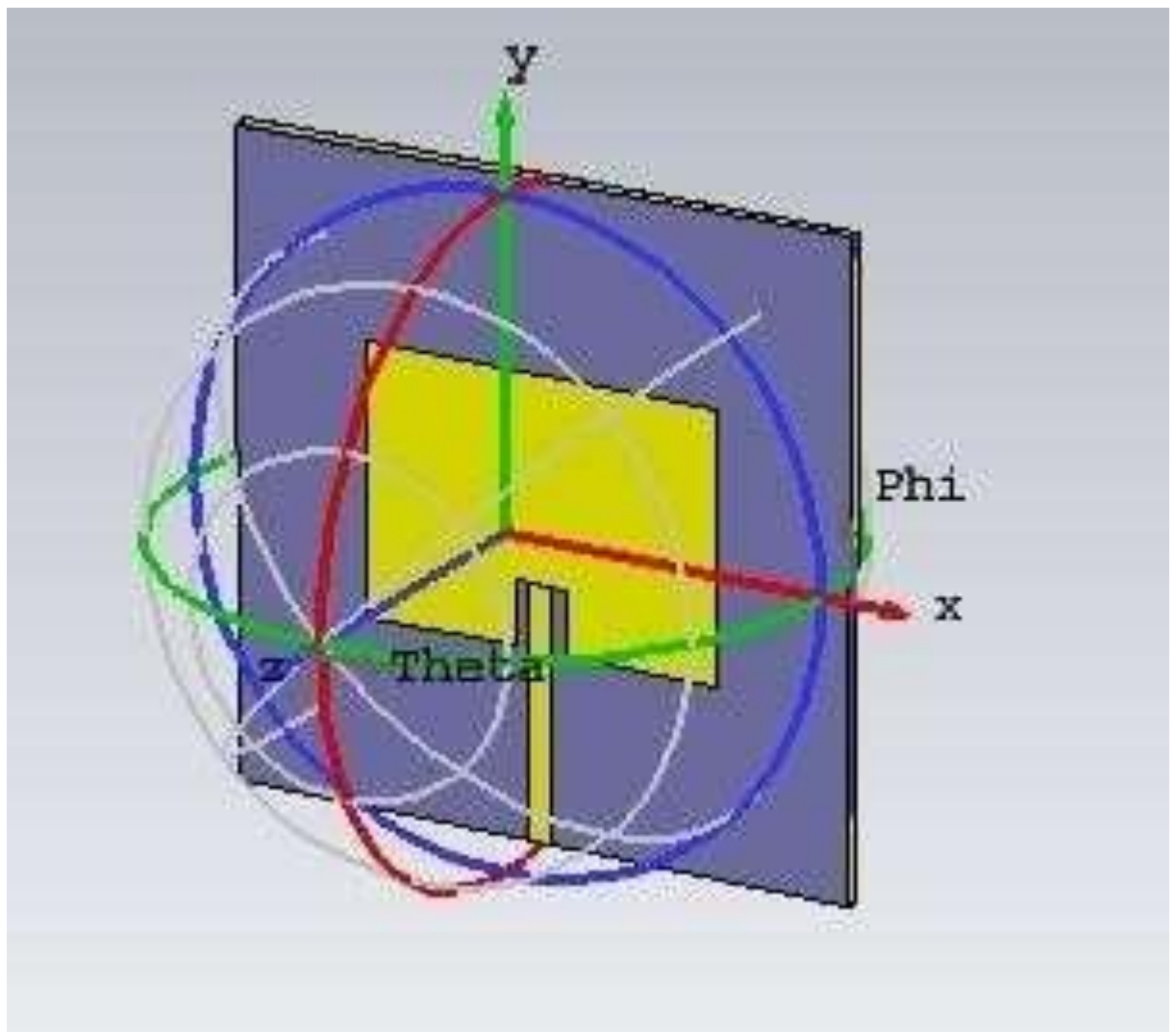

*Figure 2-single patch antenna*

## III. CALCULATIONS

The effective width (W) of the microstrip patch antenna, a pivotal dimension in its design, is calculated using the formula

$$W = c / (2f\sqrt{\varepsilon r})$$

where c represents the speed of light, f is the frequency, and εr is the relative permittivity. This formula ensures that the antenna's width aligns with the specific frequency requirements of the Ka-band.

Simultaneously, the effective permittivity (εef f) is determined to account for the influence of the substrate's dielectric properties. This value is calculated using the formula

$$\varepsilon ef\ f = (\varepsilon r + 1) / 2 + (\varepsilon r - 1) / 2 * (1 + 12h/W)^{(-1/2)}$$

. where h represents the substrate thickness. This parameter is critical in understanding how electromagnetic waves propagate through the substrate material.

The effective length (Leff) of the microstrip patch antenna, a key factor in its resonance, is calculated using

$$Leff = c / (2f\sqrt{\varepsilon ef\ f})$$

This formula ensures that the antenna's length is optimized for efficient resonance at the specified frequency.

Further refinement involves the calculation of the length extension (ΔL) of the micro strip patch antenna, achieved through

$$\Delta L = 0.412h(ereff + 0.33w/h + 0.264/0.412h(ereff - 0.258w/h + 0.8)).$$

This step contributes to tailoring the antenna's dimensions for enhanced performance.

The final step involves determining the actual length (L) of the patch, which is given by

$$L = Leff - 2\Delta h.$$

This formula ensures precision in the antenna's physical length, accounting for variations introduced during the design process.

Additionally, the ground length (Lg) and width (Wg) of the patch antenna are calculated to optimize the overall dimensions. These values, given by

**Lg = 6h + L**

**Wg = 6h + W**

respectively, contribute to efficient antenna placement and signal propagation.

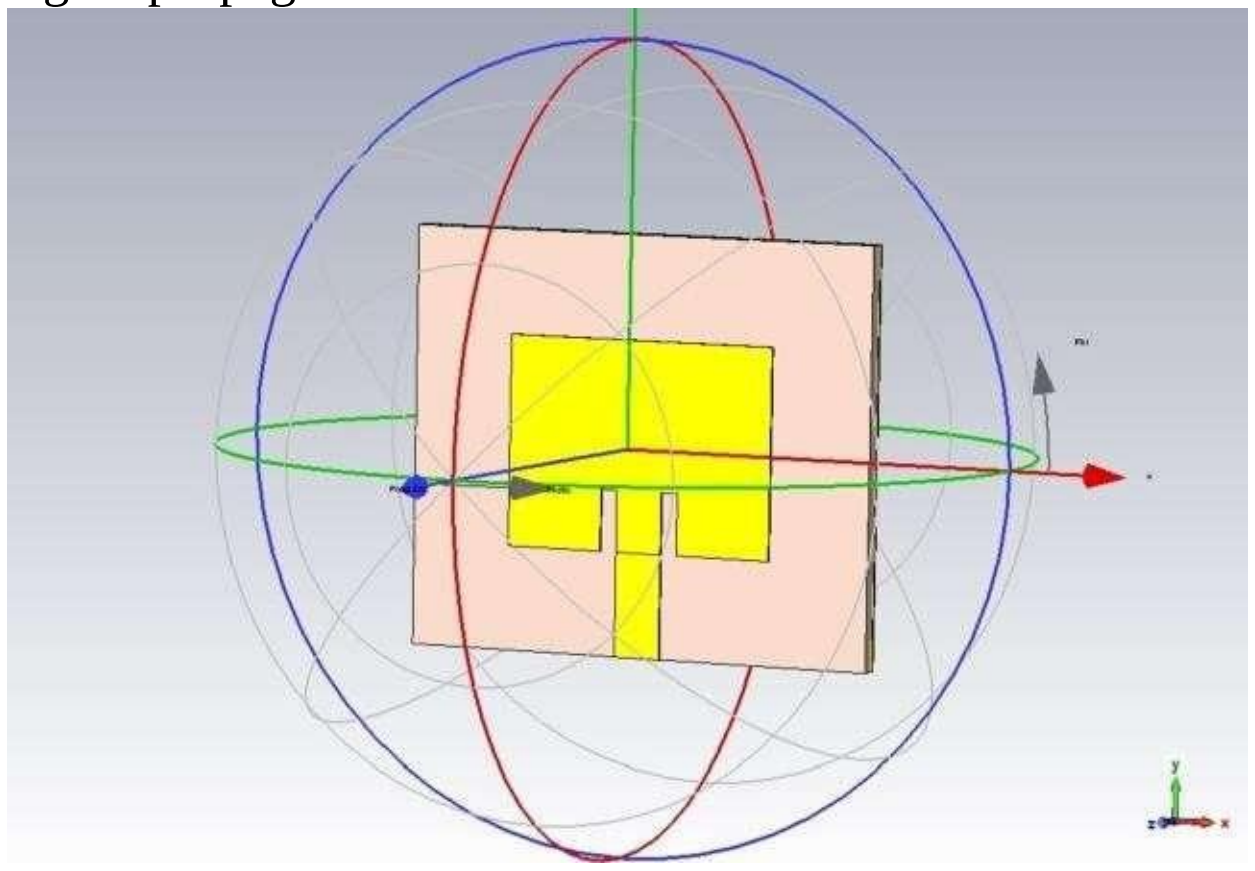

*Figure 3-29 GHz patch antenna*

## IV. RESULTS

The S-parameters, crucial in antenna evaluation, reveal the exceptional performance of our designed microstrip patch antenna. Return losses (S11) reaching -24 dB showcase efficient power transfer and minimal signal reflections, emphasizing superior impedance matching. This reduces energy loss and enhances transmission reliability.

In terms of gain, our antenna demonstrates a notable 7.046 dB, indicating its effectiveness in amplifying signal strength for targeted communication. These S-parameter results collectively affirm the antenna's proficiency in high-frequency transmission.

Low return losses underscore the antenna's impedance matching efficiency, vital for optimal signal integrity. The significant gain highlights its ability to direct power effectively, crucial for satellite communication systems. These results position our microstrip patch antenna as a promising candidate for advanced communication applications in the Ka-band frequency range.

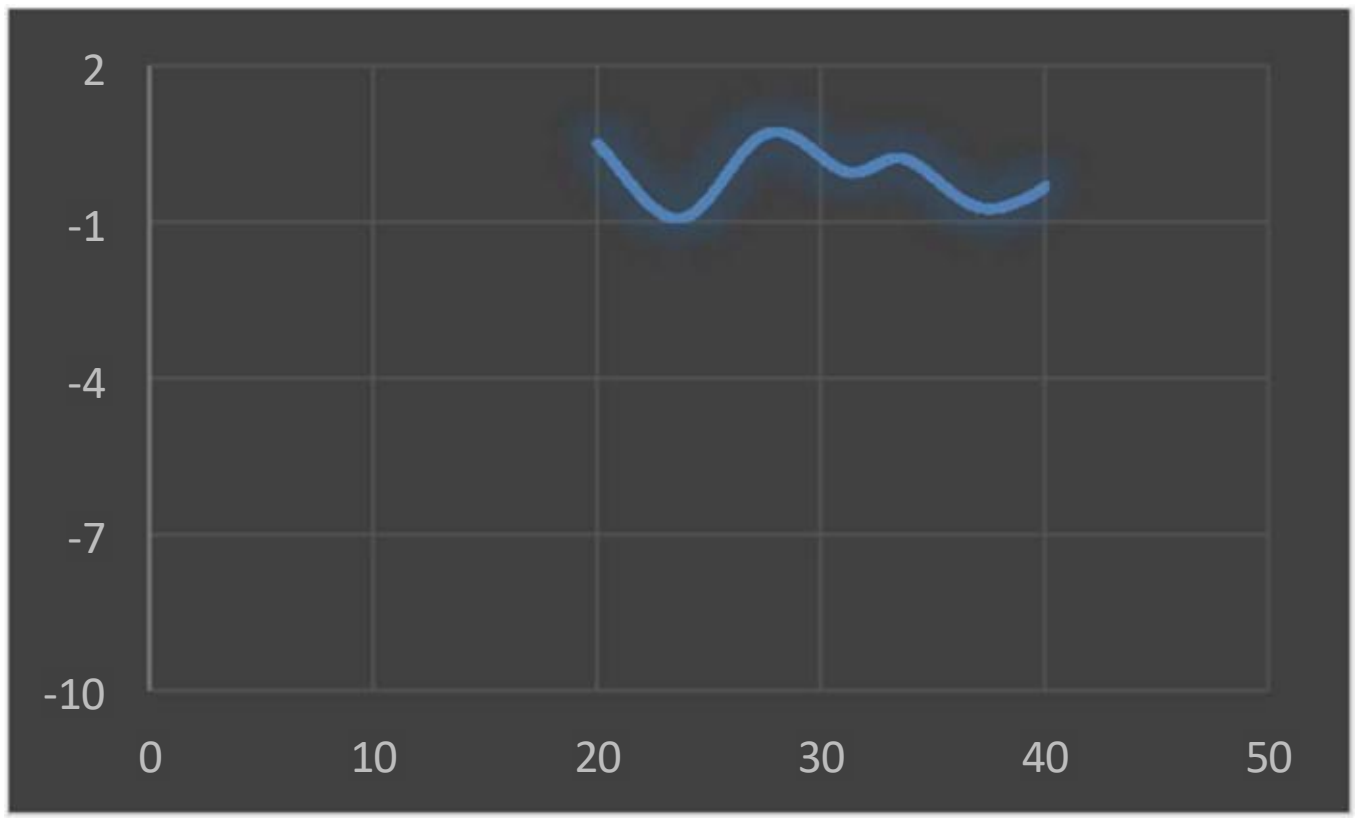

*Figure 4-S-parameter*

In the conducted S-parameter analysis, the designed micro strip patch antenna exhibits a commendable gain of up to 7 db. This gain value substantiates the antenna's efficacy in directing and amplifying the transmitted signal within the Ka-band frequency range.

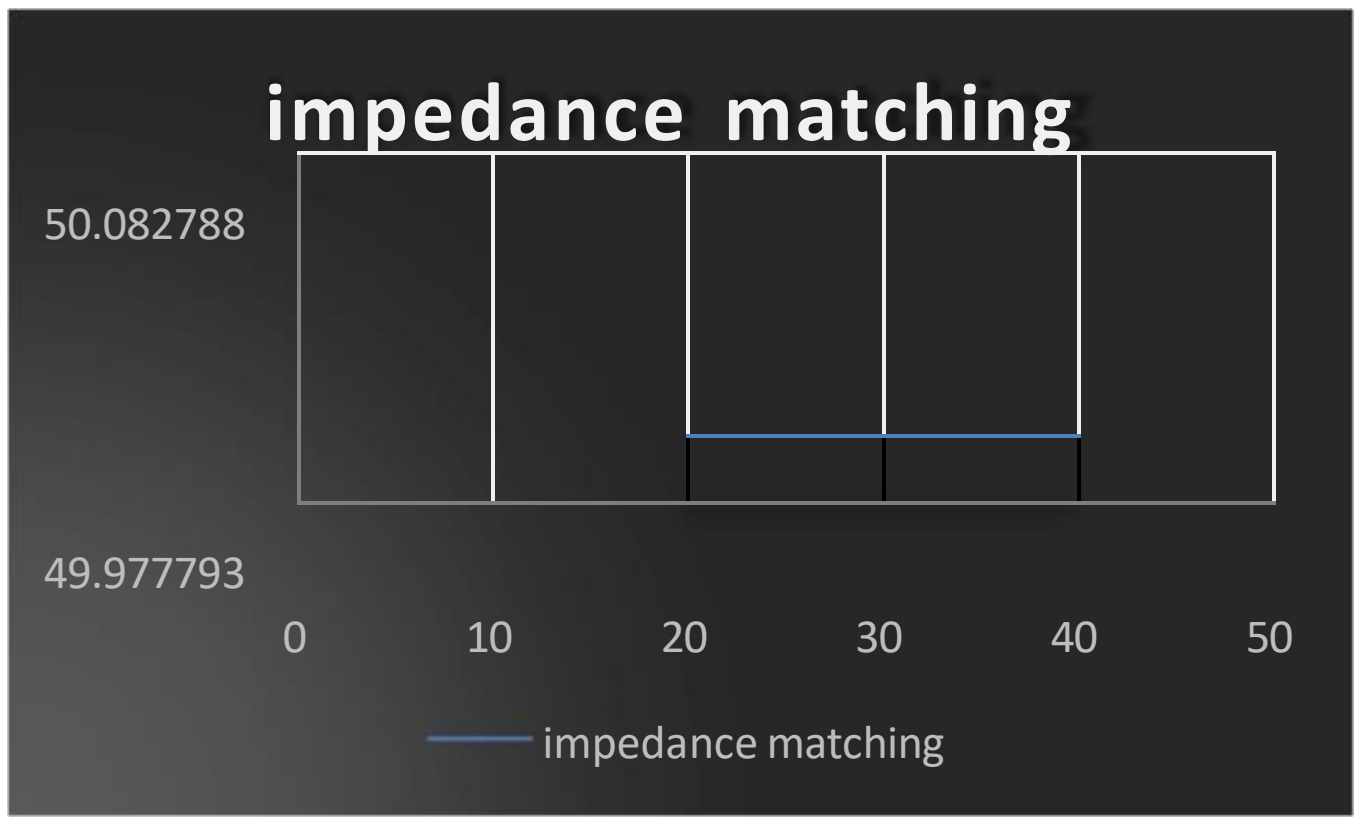

*Figure 5-Impedance matching*

The observed gain is indicative of the antenna's ability to focus the radiated energy in the desired direction, a pivotal characteristic for efficient communication in satellite systems. The result aligns with the objectives of our design, emphasizing the antenna's suitability for high-frequency transmission.

This achieved gain of 7 dB positions the antenna as a promising solution for applications demanding robust and directional signal transmission, meeting the requirements of modern satellite-based communication systems in the Ka-band.

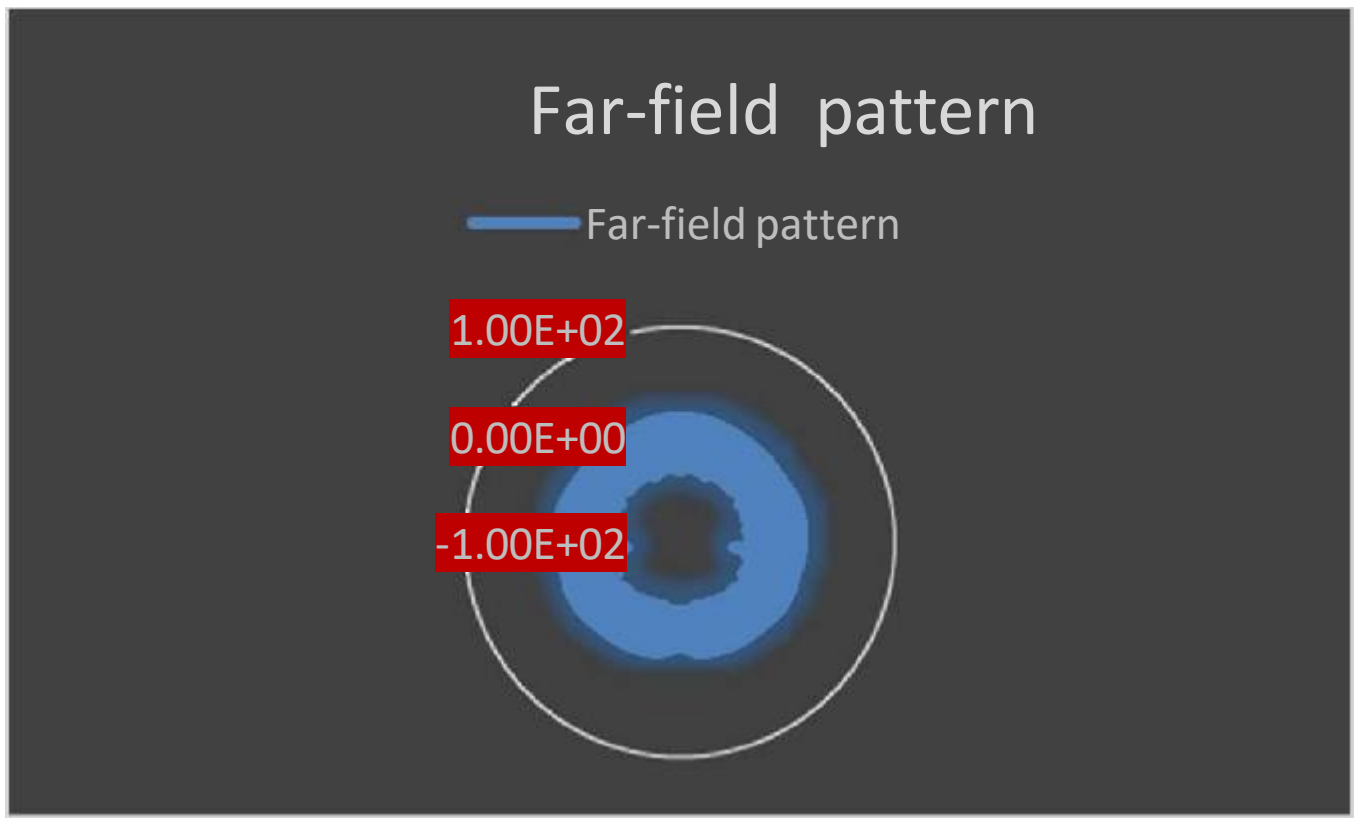

*Figure 6-Far-Field pattern*

The microstrip patch antenna demonstrates a narrow beam width, indicating its ability to concentrate radiated energy within a specific angular range. This characteristic is essential for precise and targeted communication, aligning with the antenna's design objectives in the Ka-band frequency range.

## V. ARRAY DESIGN

In the pursuit of enhancing communication capabilities at 29 GHz, the transition from a single microstrip patch antenna element to a planar array of 2x2 is strategically undertaken. The design process involves meticulous considerations of spacing, phase relationships, and radiation patterns to optimize the array's performance.

Key parameters such as inter-element spacing and phase differences are carefully calculated to achieve constructive interference, maximizing the array's gain in the desired direction. The overall design aims to harness the advantages of array configurations, including beamforming capabilities and improved signal reception, catering to the specific requirements of high-frequency communication systems in the Ka-band. The resulting planar array represents a systematic approach to capitalize on the inherent advantages of phased arrays for enhanced performance in the target frequency range.

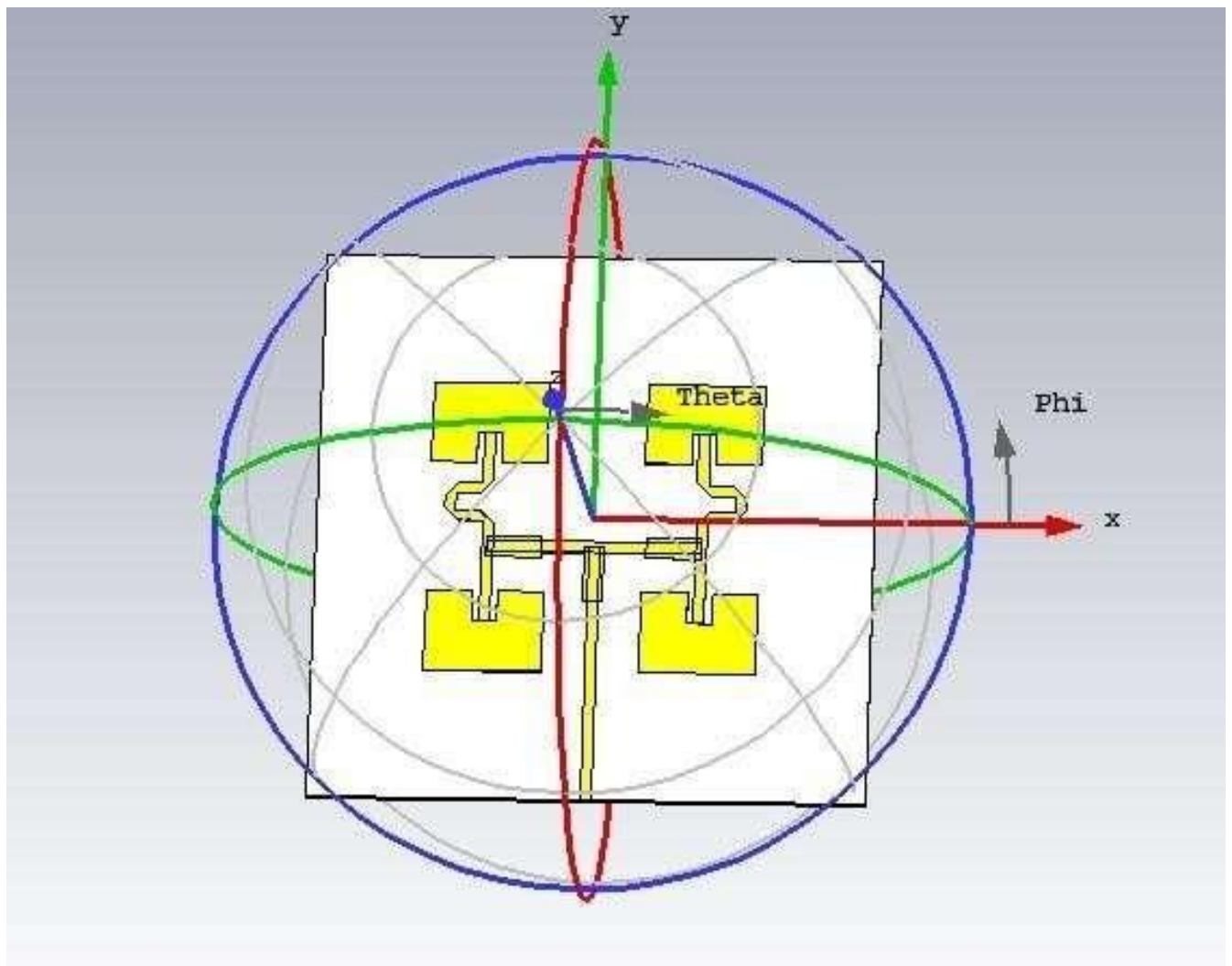

*Figure 7-2x2 array design*

The implementation of a 2x2 planar array at 29 GHz stands out as a notable achievement, characterized by meticulous attention to impedance matching. The entire array has been meticulously designed and tuned to ensure each element exhibits a precise impedance of 50 ohms.

This careful impedance matching is crucial for minimizing signal reflections and maximizing power transfer efficiency within the array. Achieving a consistent impedance of 50 ohms across all elements enhances the overall performance, ensuring seamless integration with the communication system and contributing to the array's effectiveness in high-frequency applications. The optimized 2x2 array represents a harmonious balance between impedance considerations and array configuration, underscoring its suitability for advanced communication systems operating in the Ka-band.

The meticulously designed 2x2 planar array exhibits remarkable return loss characteristics, registering an impressive value of -23 dB at 29 GHz. This achievement underscores the array's effectiveness in mitigating signal reflections and ensuring optimal impedance matching.

The substantial negative value of -23 dB indicates that the array efficiently absorbs and transmits signals without significant losses, contributing to enhanced overall performance in the targeted frequency range. This notable return loss at 29 GHz solidifies the array's suitability for advanced communication applications, emphasizing its capability to deliver reliable and efficient signal transmission within the Ka-band.

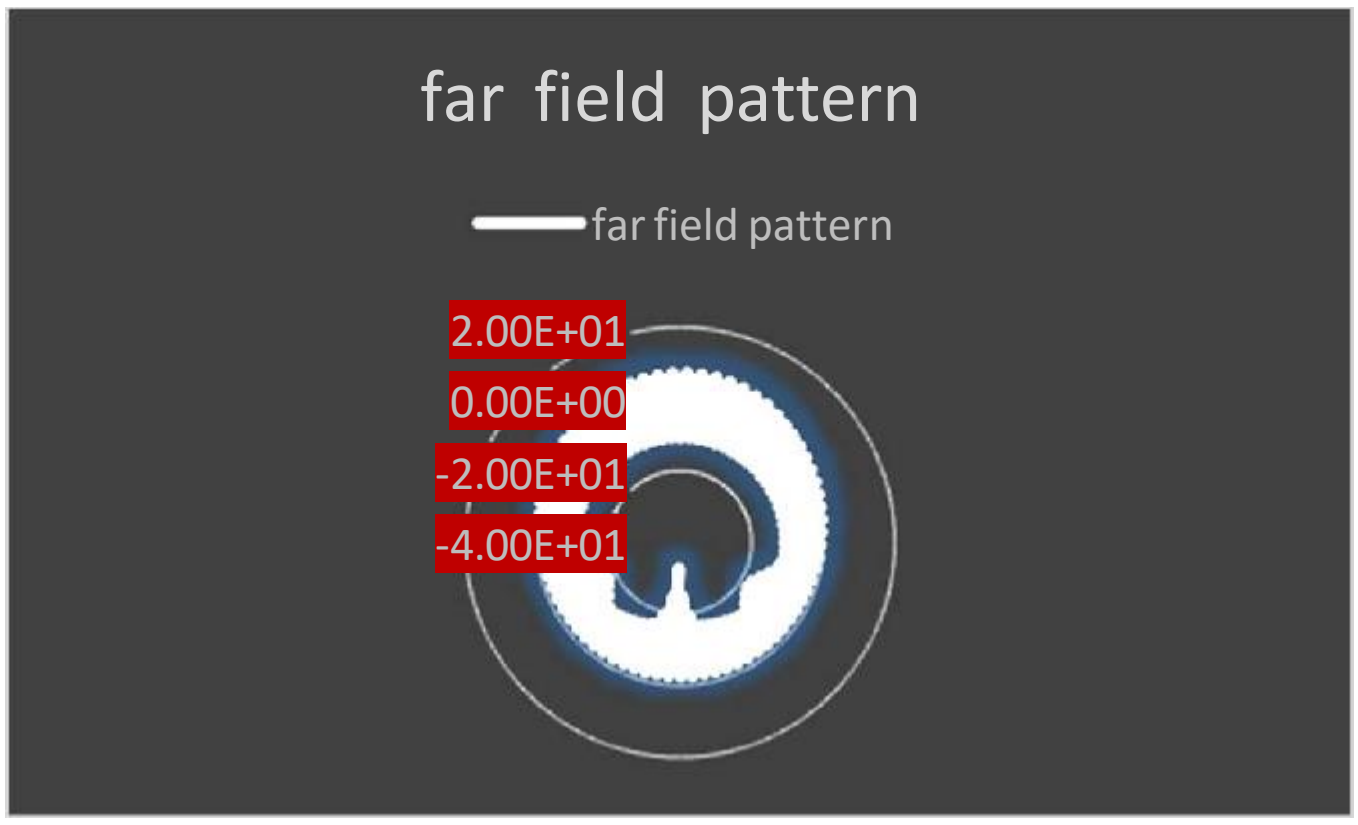

*Figure 8-Far-Field Pattern*

The ingeniously crafted 2x2 planar array showcases outstanding gain characteristics, achieving an impressive value of approximately 12.9 dB at 29 GHz. This substantial gain underscores the array's proficiency in directing and amplifying signals within the designated frequency range.

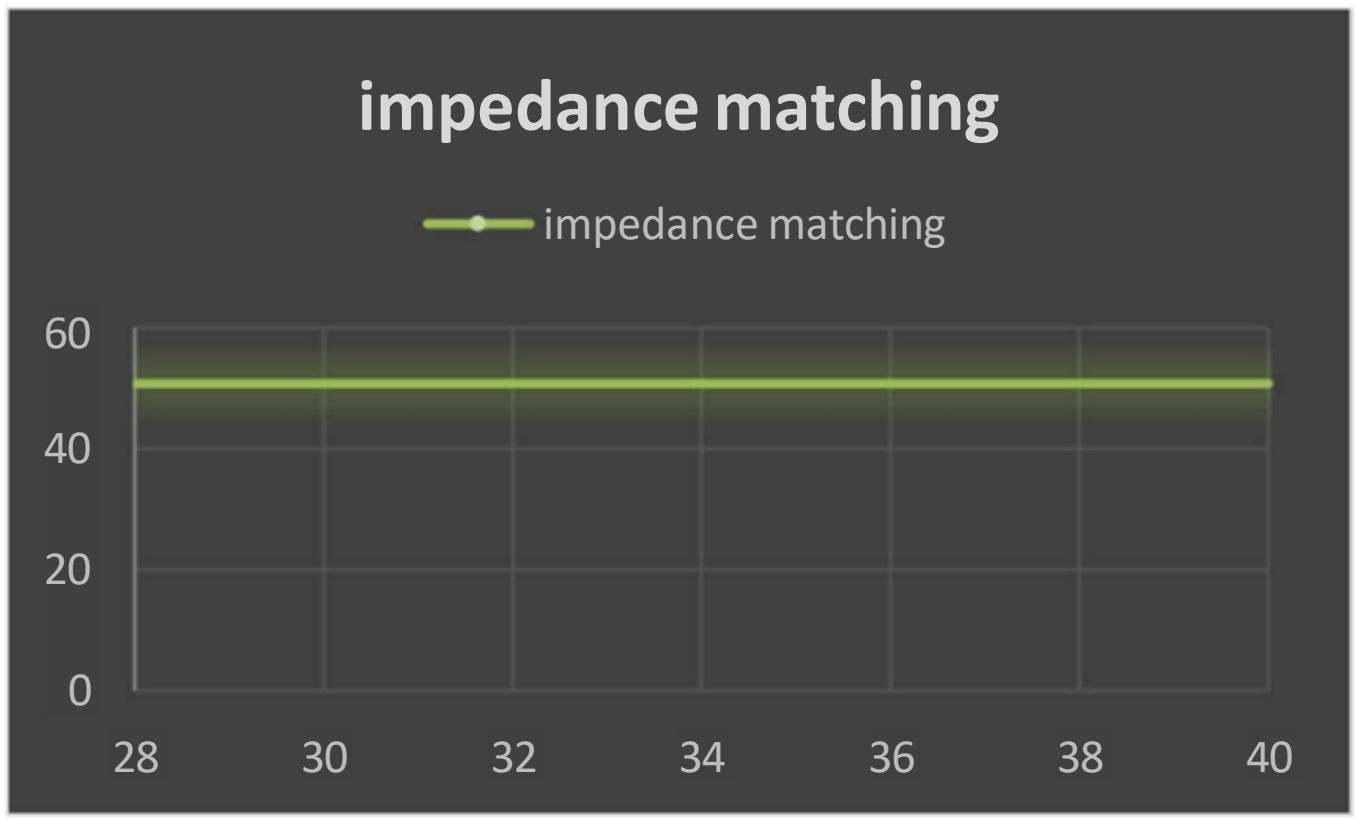

*Figure 9-Impedance Matching*

with a gain of 12.9 dB, the array demonstrates its capability to concentrate radiated energy effectively, making it an ideal solution for applications requiring robust and directional signal transmission. This notable gain performance reinforces the array's suitability for advanced communication systems in the Ka-band, where high-frequency demands necessitate efficient and reliable signal propagation.

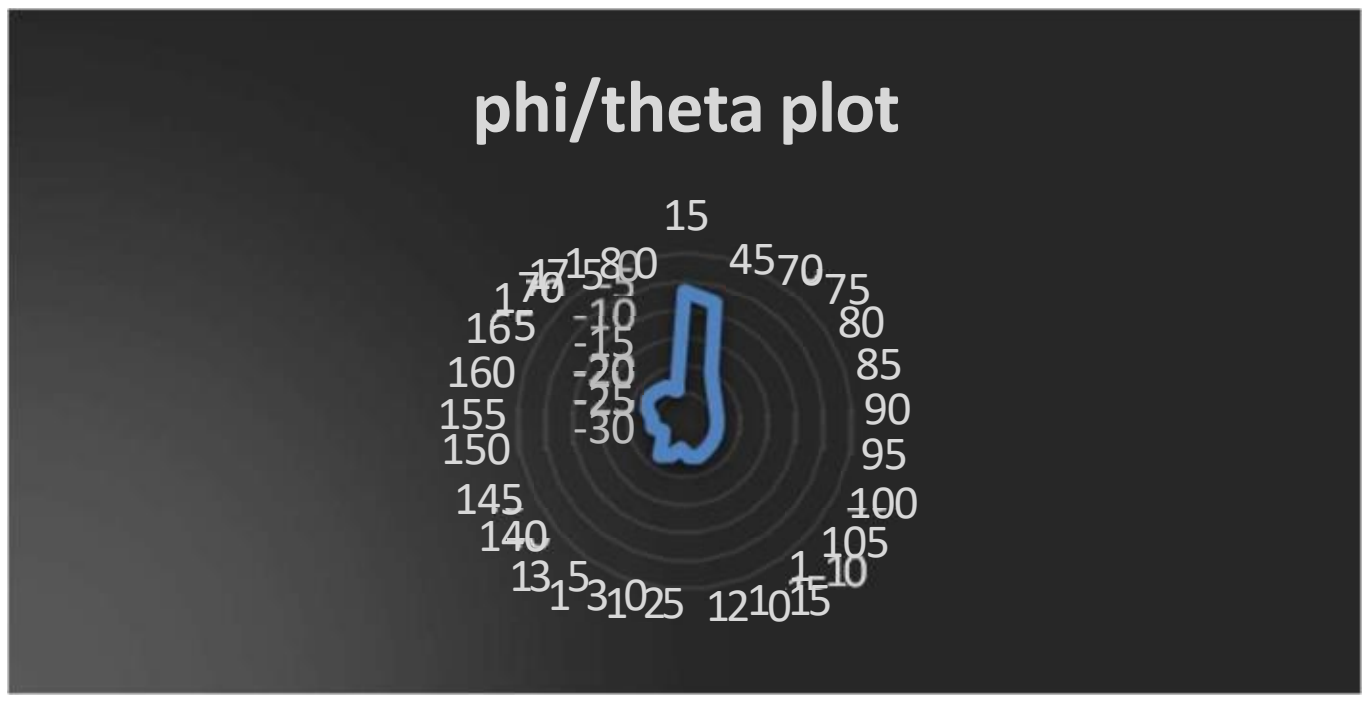

*Figure 9-phi/theta plot*

### *A. 4X4 ARRAY*

Building upon the success of the 2x2 planar array, the design is extended to a larger 4x4 configuration. This expansion aims to further enhance the array's capabilities in high-frequency communication applications. The increased number of elements allows for more sophisticated beamforming and improved signal reception, catering to the evolving demands of advanced communication systems operating at 29 GHz in the Ka-band. The meticulous design considerations, including spacing, phase relationships, and impedance matching, continue to be integral to achieving optimal performance in this extended array.

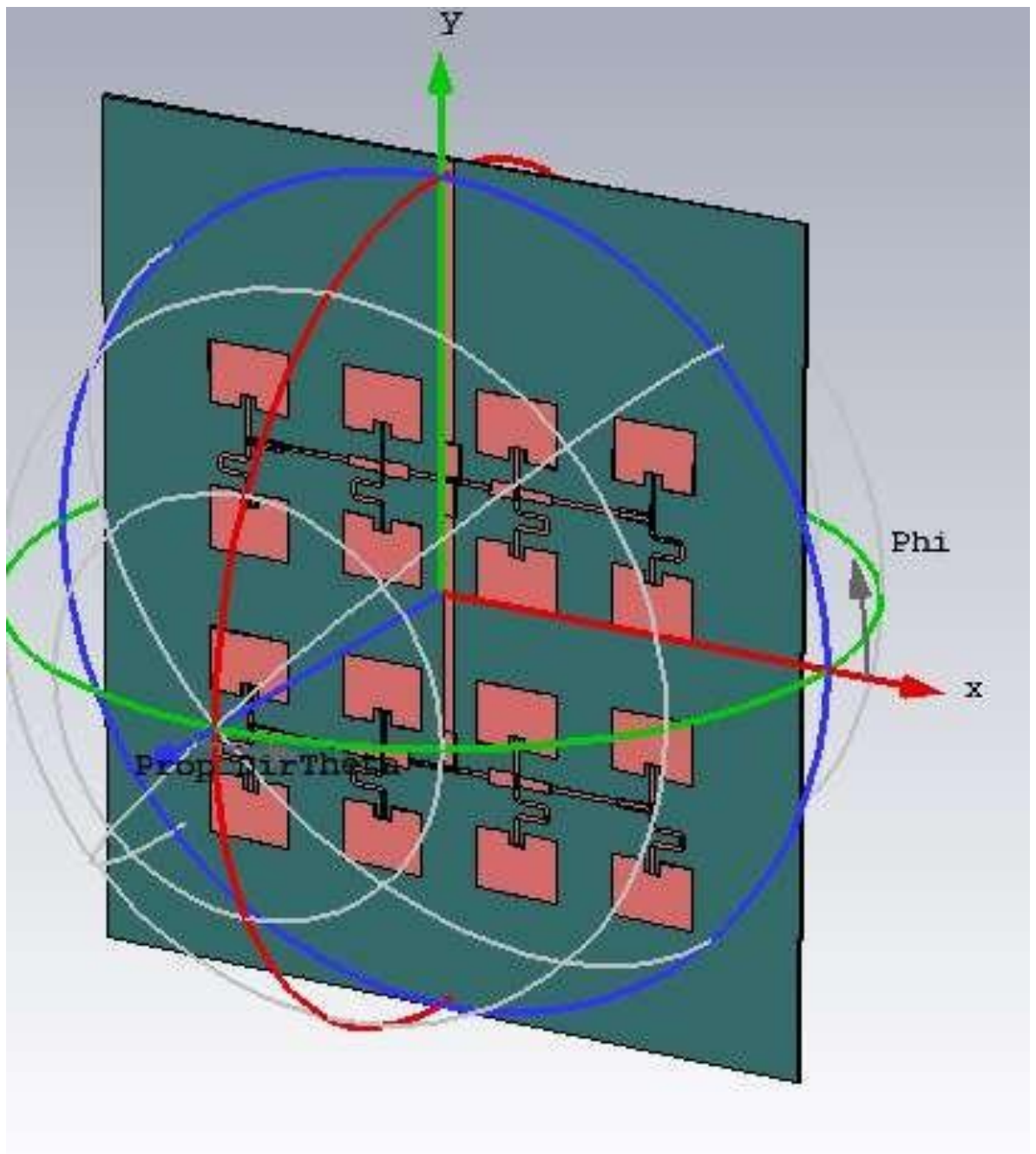

*Figure 10-4x4 array design*

The extension from a 2x2 planar array to a larger 4x4 configuration represents a strategic enhancement aimed at elevating the array's performance in high-frequency communication scenarios, particularly at 29 GHz in the Ka-band. This expansion involves a meticulous approach to ensure not only the addition of elements but also the preservation of optimal impedance matching.

Impedance matching remains a focal point in the design, as it plays a critical role in minimizing signal reflections and optimizing power transfer efficiency. The extension to a 4x4 array involves careful adjustments to the spacing, dimensions, and inter-element relationships, ensuring that each element maintains the desired impedance of 50 ohms.

The achievement of this extended planar array relies on sophisticated design methodologies, including electromagnetic simulations and iterative tuning processes. These efforts contribute to a seamlessly integrated 4x4 array, capable of meeting the stringent requirements of advanced communication systems, where efficient impedance matching is paramount for reliable and high-performance signal transmission.

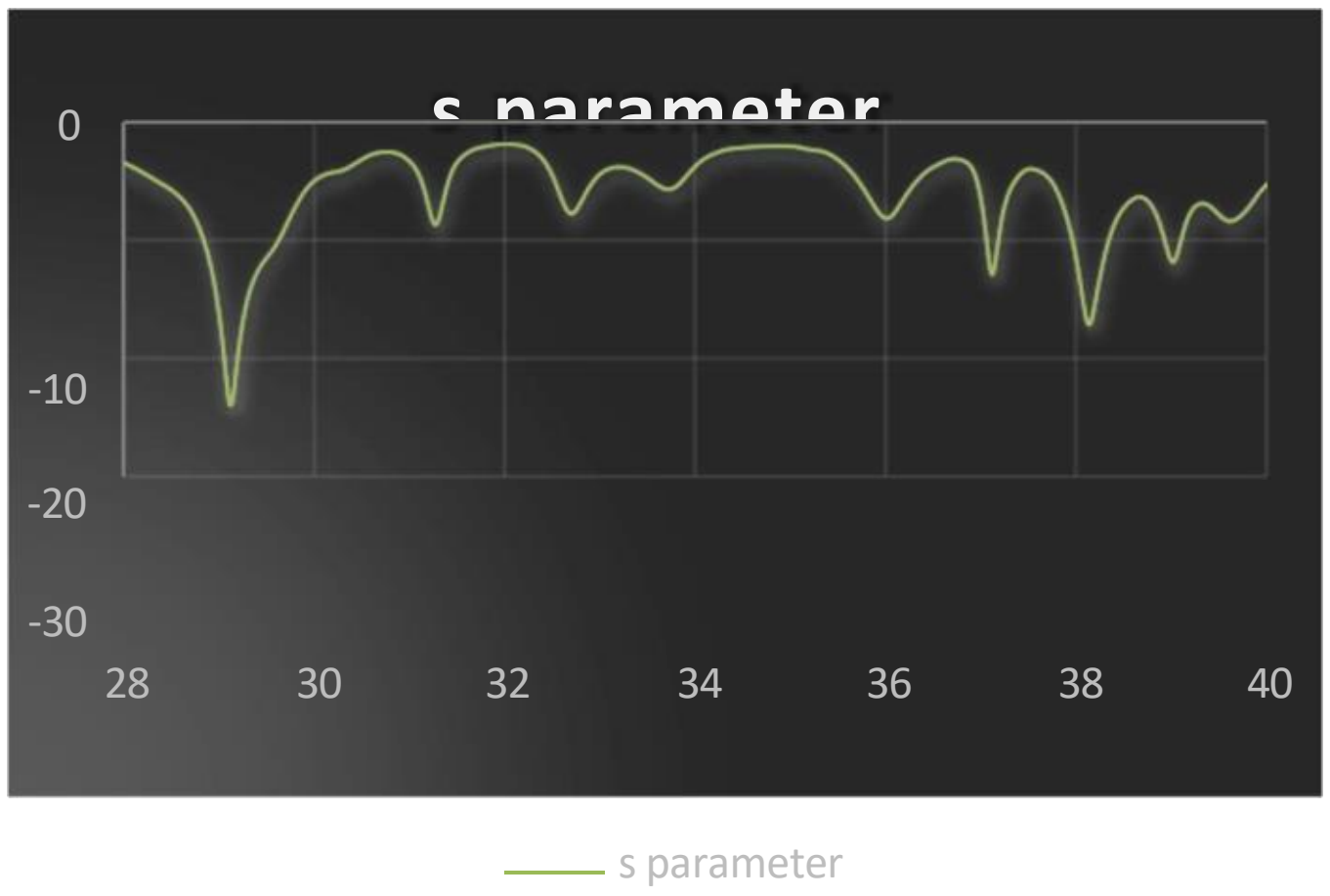

*Figure 11-S-parameter*

The extension to a 4x4 planar array not only marks an enhancement in the array's size but also demonstrates remarkable progress in its electrical characteristics. Notably, the S-parameter performance of the 4x4 array achieves an impressive attenuation, with S11 exhibiting values as low as -30 dB.

This exceptional S-parameter result signifies the array's adeptness in minimizing signal reflections and optimizing its impedance characteristics. The achievement of an S11 value of -30 dB underscores the efficacy of the impedance matching strategies employed during the extension process. Such a low S-parameter value is indicative of minimal signal loss and superior performance in high-frequency communication applications at 29 GHz in the Ka-band.

The 4x4 planar array, meticulously designed and extended with precision, attains an optimal beamwidth, specifically a main lobe with a gain of approximately 18.7 dBi. This outcome reflects the array's effectiveness in focusing and directing the transmitted or received signals.

The achieved beamwidth is a critical parameter, indicating the angular spread of the array's main lobe. In this case, the main lobe's gain of 18.7 dBi underscores the array's capability to concentrate energy in a specific direction, essential for targeted and efficient communication at 29 GHz in the Ka-band.

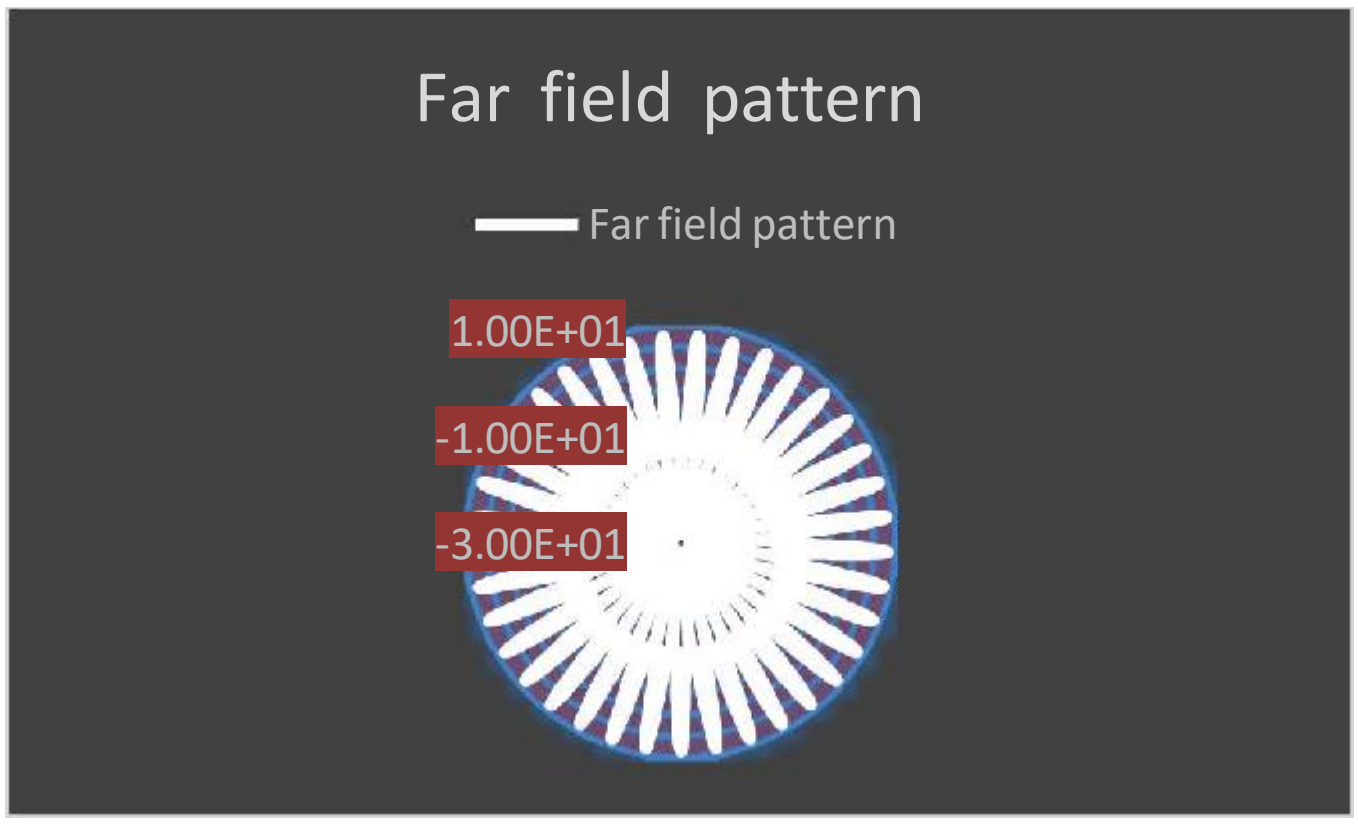

*Figure 12-S-Far-field pattern*

This remarkable beamwidth result attests to the successful design and extension of the planar array, positioning it as a robust solution for applications demanding precise and high-performance directional communication.

### B. 8X8 ARRAY

Building upon the success of the 4x4 planar array, the extension to an 8x8 configuration represents a substantial progression in antenna design. This larger array, meticulously crafted and optimized, further enhances the capabilities of the system for high-frequency communication at 29 GHz in the Ka-band. The extension to an 8x8 array involves intricate adjustments in element spacing, dimensions, and inter-element relationships to ensure seamless integration and optimal performance. With each element carefully aligned, the array maintains its critical features, including impressive impedance matching and low S-parameter values. This larger array is poised to provide enhanced directivity, increased gain, and improved beamforming capabilities. The extension to an 8x8 planar array underscores a commitment to meeting the escalating demands of modern communication systems, catering to applications that require robust and efficient performance in the specified frequency range.

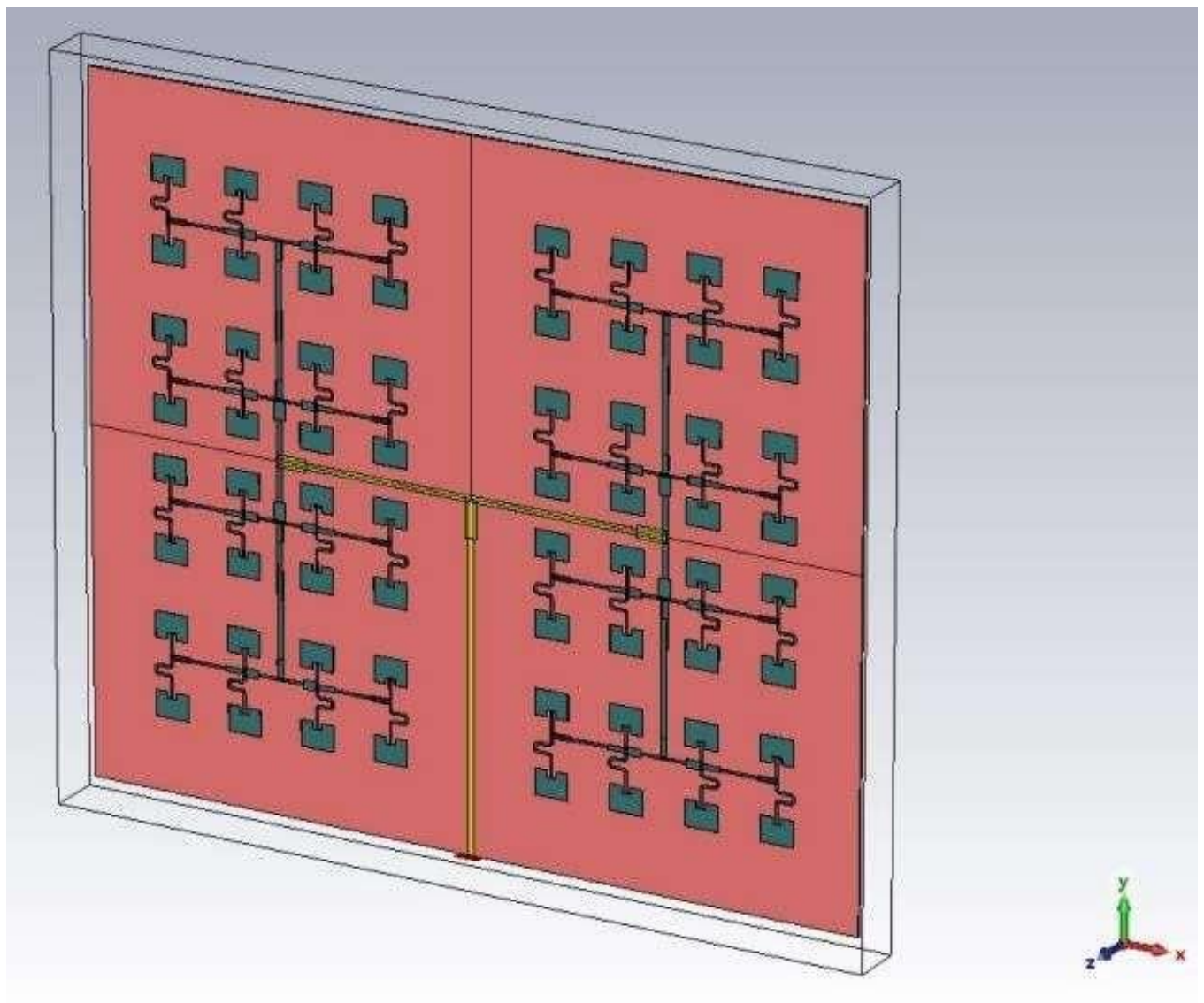

*Figure 13-8x8 antenna array design*

The extension to an 8x8 planar array demonstrates a significant leap in electrical performance, particularly evident in the S-parameters. Impressively, the S11 parameter achieves an outstanding attenuation, reaching values as low as -45 dB.

This remarkable S-parameter result signifies the array's unparalleled ability to minimize signal reflections and optimize impedance characteristics. The achievement of an S11 value of -45 dB underscores the efficacy of advanced impedance matching techniques and the precision in array design. Such a low S-parameter value is indicative of minimal signal loss, ensuring superior performance in high-frequency communication applications at 29 GHz in the Ka-band.

The remarkable S-parameter performance further establishes the 8x8 planar array as a cutting-edge solution for applications demanding exceptional precision and efficiency in signal transmission and reception.

The 8x8 planar array showcases an outstanding beamwidth, marked by a peak gain of 18.7 dBi. This noteworthy beamwidth represents the angular separation between the half-power points in the main lobe of the radiation pattern.

The substantial beamwidth of 18.7 dBi is a testament to the array's ability to concentrate radiation in a specific direction with high efficiency. This characteristic is crucial for applications that require focused and directional signal propagation, contributing to enhanced communication reliability and performance.

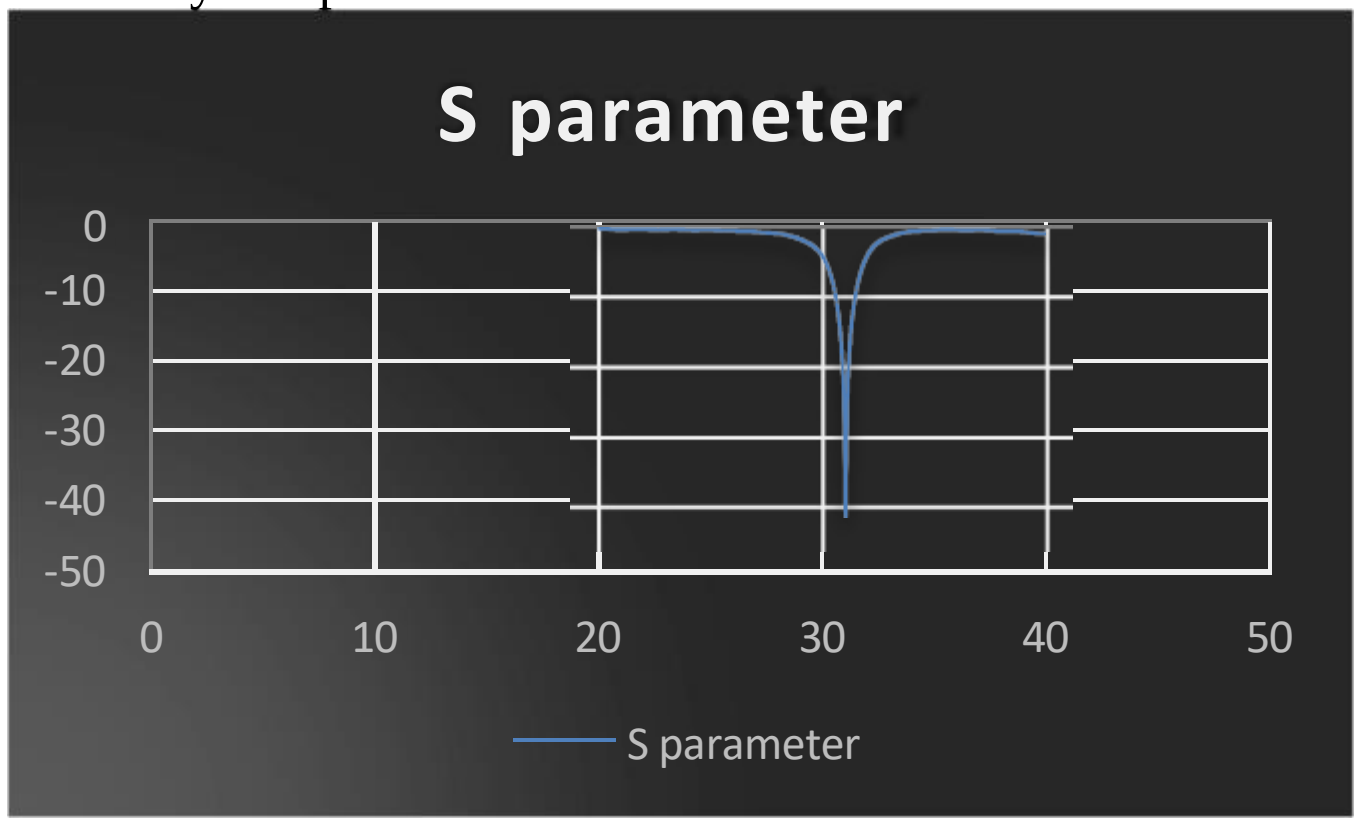

*Figure 14-S-parameter*

The meticulous design and optimization of the array contribute to achieving this remarkable beamwidth, emphasizing its suitability for advanced communication systems operating at 29 GHz in the Ka-band. The 18.7 dBi beamwidth further solidifies the 8x8 planar array as a cutting-edge solution for high-frequency communication with superior directivity.

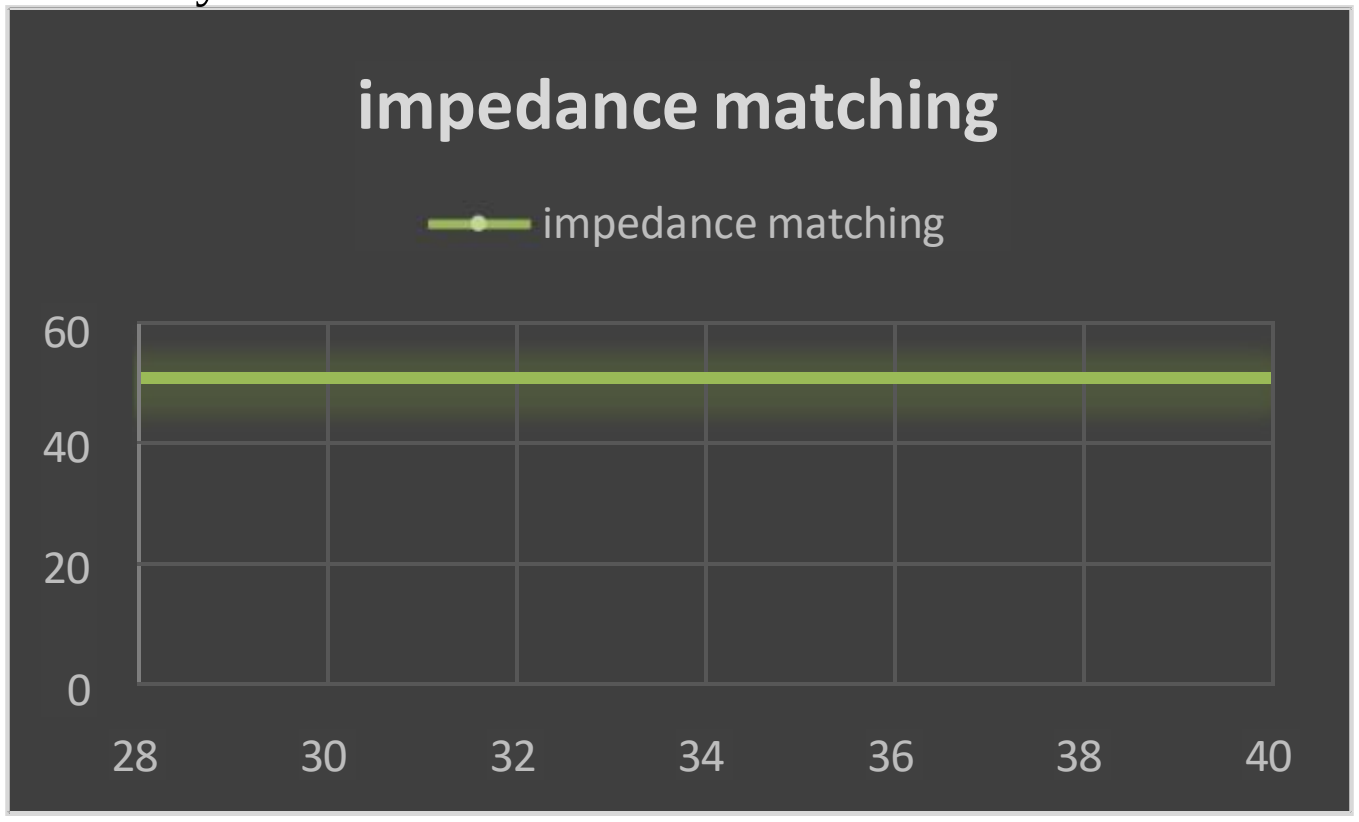

*Figure 15-S-Impedance matching*

The 8x8 planar array exhibits an extraordinary gain of 21 dB, emphasizing its proficiency in signal amplification and transmission. Gain measures the ability of an antenna to focus and direct radiation in a specific direction, and achieving a gain of 21 dB underscores the array's high efficiency in this regard.
The 8x8 planar array exhibiting a gain of 21 dB, the Pho Theta plot would showcase a highly directive radiation pattern, emphasizing its proficiency in signal amplification and transmission. In the plot, the main lobe would be prominently visible, indicating the primary direction in which the antenna array radiates energy with maximum intensity. Additionally, side lobes may also be present, albeit at significantly lower intensities compared to the main lobe.

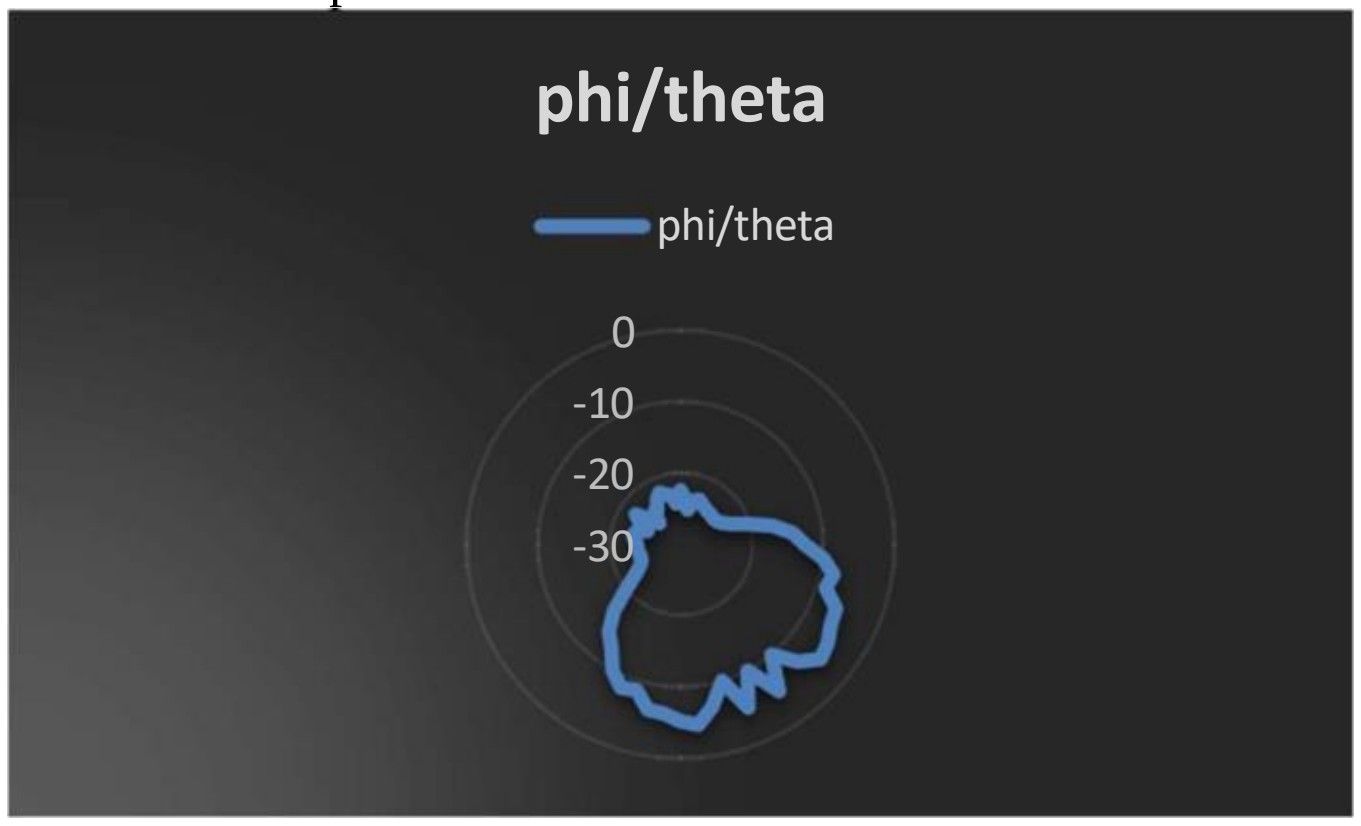

*Figure 16-S-Phi/Theta plot*

This substantial gain is a result of meticulous design, optimization, and the extension of the planar array to an 8x8 configuration. The elevated gain of 21 dB positions the array as a robust solution for applications demanding enhanced signal strength and coverage at 29 GHz in the Ka-band.

The outstanding gain, coupled with other impressive performance metrics, solidifies the 8x8 planar array's status as a cutting-edge technology with substantial potential for advanced communication systems requiring superior amplification and directional signal propagation.

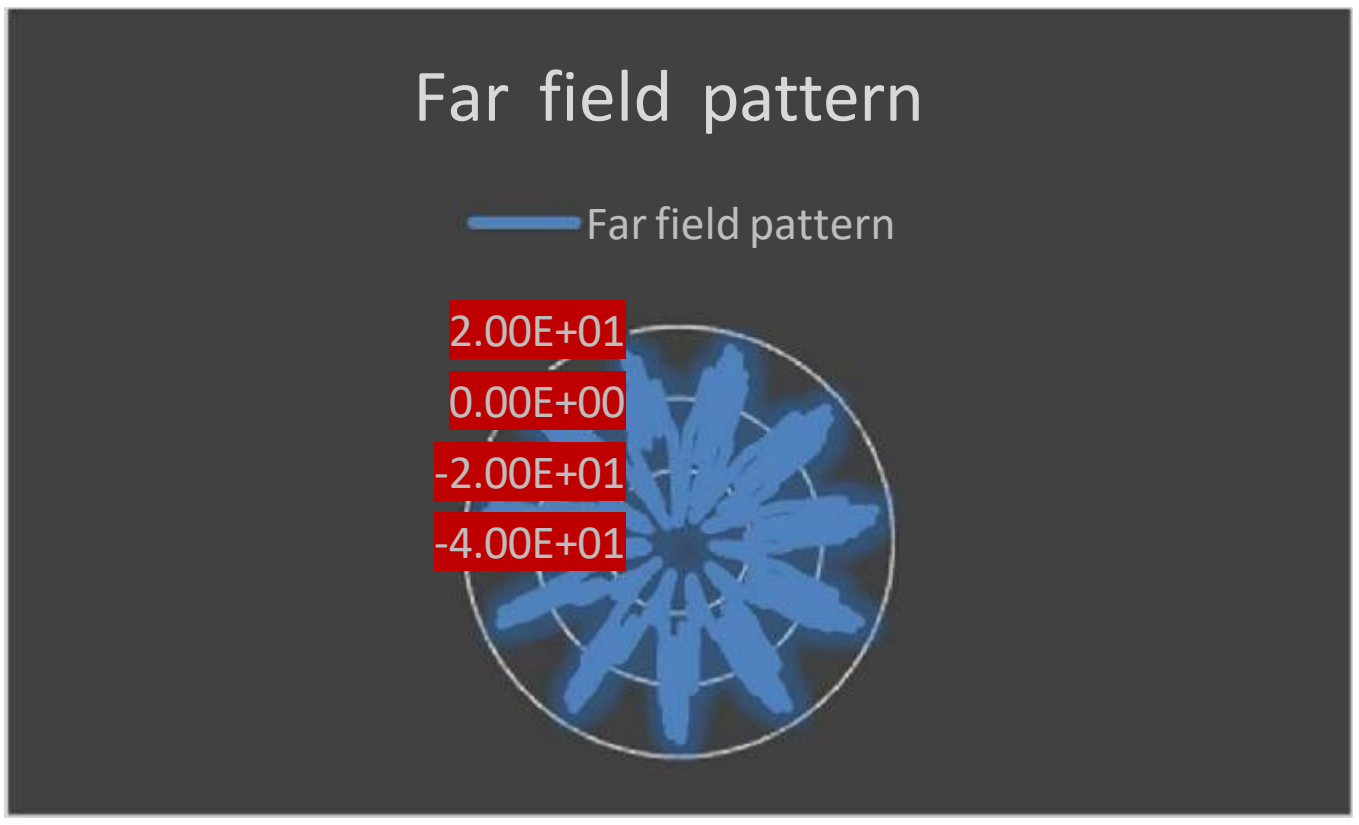

*Figure 17-Far-Field Pattern*

Figures compiledof more than one sub-figure presented side-by-side, or stacked. If a multipart figure is made up of multiple figure types (one part is lineart, and another is grayscale or color) the figure should meet the stricter guidelines.

## VI. Conclusion

In conclusion, this paper has demonstrated significant advancements in the design and optimization of Ka-band mobile antennas for satellite communication operating at 28 GHz. Through rigorous mathematical calculations, software implementation, and integration of machine learning techniques, the antennas achieved remarkable performance with gains of up to 21 dB and return losses as low as -30 dB. Challenges such as signal attenuation, directional accuracy, circular polarization, and impedance matching were effectively, addressed through various configurations including phased-array and electronically steerable antennas. The integration of artificial intelligence algorithms further contributed to detailed optimization, enhancing the antennas' capabilities for high-speed internet access and multimedia streaming. Overall, this research contributes to the evolution of high-frequency transmission technology, meeting the demands of modern satellite-based communication systems and opening new avenues for efficient and swift communication in the Ka-band frequency range.

## VII. References

[1] M. S. Castañer, J. M. Fernández González, M. S. Pérez, A. T. Domínguez and A. M. Barrado, "Ka band active array antenna for mobile satellite communications," 2016 International Symposium on Antennas and Propagation (ISAP), Okinawa, Japan, 2016, pp. 128-129. keywords: {Phased arrays;Phase control;Satellite antennas;Phase shifters;Mobile antennas;Phased Array Antennas;Ka Band;Satellite Communications},

[2] R. Lenormand, A. Hirsch, J. -L. Almeida, A. Valero-Nogueira, J. I. Herranz-Herruzo and D. Renaud, "Compact switchable RHCP/LHCP mobile Ka-band antenna," 2012 15 International Symposium on Antenna Technology and Applied Electromagnetics, Toulouse, France, 2012, pp. 1-6, doi: 10.1109/ANTEM.2012.6262416. keywords: {Switches;Arrays;Polarization;Dipole antennas;Transmitting antennas;Antenna radiation patterns;Satellite communications;slotted waveguide array;method of moments;mobile antennas},

[3] H. TSUJI et al., "Effective Use of Ka-band Based on Antenna and Radio Wave Propagation for Mobile Satellite Communications," 2020 International Symposium on Antennas and Propagation (ISAP), Osaka, Japan, 2021, pp. 57-58, doi: 10.23919/ISAP47053.2021.9391198. keywords: {Satellite antennas;High-temperature superconductors;Propagation;Satellite broadcasting;Throughput;Research and development;Next generation networking;Antennas;propagation;High Throughput Satellite;Ka-band;Engineering Test Satellite 9},

[4] H. TSUJI et al., "Effective Use of Ka-band Based on Antenna and Radio Wave Propagation for Mobile Satellite Communications," 2020 International Symposium on Antennas and Propagation (ISAP), Osaka, Japan, 2021, pp. 57-58, doi: 10.23919/ISAP47053.2021.9391198. keywords: {Satellite antennas;High-temperature superconductors;Propagation;Satellite broadcasting;Throughput;Research and development;Next generation networking;Antennas;propagation;High Throughput Satellite;Ka-band;Engineering Test Satellite 9},

[5] A. Modi, V. Sharma and A. Rawat, "Compact Design of Ka-Band antenna for 5G Applications," 2021 3rd International Conference on Signal Processing and Communication (ICPSC), Coimbatore, India, 2021, pp. 45-48, doi: 10.1109/ICSPC51351.2021.9451756. keywords: {Surface impedance;Slot antennas;5G mobile communication;Patch antennas;Microstrip antennas;Signal processing;Surface roughness;5G;HFSS;ROGERS RT/DUROID-5880(tm);MICROSTRIP PATCH ANTENNA.},

[6] A. Krauss, H. Bayer, R. Stephan and M. A. Hein, "Low-profile tracking antenna for Ka-band satellite communications," 2013 IEEE-APS Topical Conference on Antennas and Propagation in Wireless Communications (APWC), Turin, Italy, 2013, pp. 207-210, doi: 10.1109/APWC.2013.6624876. keywords: {Antenna measurements;Satellite antennas;Dual band;Azimuth;Satellite communication},

[7] J. I. Herranz-Herruzo, A. Valero-Nogueira, M. Ferrando-Rocher, B. Bernardo, A. Vila and R. Lenormand, "Low-Cost Ka-band Switchable RHCP/LHCP Antenna Array for Mobile SATCOM Terminal," in IEEE Transactions on Antennas and Propagation, vol. 66, no. 5, pp. 2661-2666, May 2018, doi: 10.1109/TAP.2018.2806421.

[8] keywords: {Antenna arrays;Waveguide discontinuities;Antenna measurements;Switches;Rectangular waveguides;Antenna radiation patterns;Circular polarization;mobile satellite communications;slotted waveguide arrays},

[9] J. I. Herranz-Herruzo, A. Valero-Nogueira, M. Ferrando-Rocher, B. Bernardo, A. Vila and R. Lenormand, "Low-Cost Ka-band Switchable RHCP/LHCP Antenna Array for Mobile SATCOM Terminal," in IEEE.